\documentclass[amsmath,amssymb,prl]{revtex4}
\usepackage{graphicx}

\begin{document}

\title{Asymptotic analysis of the model for distribution of high-tax payers}

\author{Hiroshi Yamamoto}
\email{hirobrd@abox.so-net.ne.jp}
\author{Toshiya Ohtsuki}
\affiliation{Division of Science, International College of Arts and Science, \\
Yokohama City University, 22-2, Seto, Kanazawa-ku, Yokohama, 236-0027}

\author{Akihiro Fujihara and Satoshi Tanimoto}
\affiliation{Graduate School of Integrated Science, Yokohama City University, 
22-2, Seto, Kanazawa-ku, Yokohama, 236-0027}

\author{Keizo Yamamoto}
\affiliation{Faculty of Engineering, Setsunan University, 17-8 Ikeda-Nakamachi,
Neyagawa, Osaka 572-8508}

\author{Sasuke Miyazima}
\affiliation{Department of Natural Science, Chubu University, 1200 Matsumoto, Kasugai,
Aichi 487-8501}

\date{\today}

\begin{abstract}
The z-transform technique is used to investigate the model for distribution of 
high-tax payers, which is proposed by two of the authors (K. Y and S. M)
and others\cite{Kyamamoto, Kyamamoto2, Kyamamoto3}. Our analysis
shows an asymptotic power-law of this model with the exponent -5/2 when a 
total ``mass'' has a certain critical value. Below the critical value, the system 
exhibits an ordinary critical behavior, and scaling relations hold. Above the 
threshold, numerical simulations show that a power-law distribution coexists 
with a huge ``monopolized'' member. 
It is argued that these behaviors are observed universally 
in conserved aggregation processes, by analizing an extended model.
\end{abstract}


\maketitle

\vspace{0.5cm}


\vspace{1cm}

\noindent
{\large\bf 1. Introduction}

Some years ago, two of the authors (K. Y and S. M) {\it et al}
\cite{Kyamamoto,Kyamamoto2,Kyamamoto3} 
proposed a simple model for distribution of high-tax payers. They considered 
a 2-body interaction system where the winner takes all competing money 
when the competition occurs. Their model is regarded as a simple aggregation 
system, which is widely discovered in various systems in nature and, naturally,
has been attracted considerable interest by many researchers\cite{ley, Schmelzer}
. Their simulation gives the universal power-law distribution, and shows 
a good fit with the date in Japan and the United States' CEO's.
However they could not succeed in theoretical derivation of obtained results.

We analyze their model mathematically, and derive an asymptotic expression
of the probability distribution functions. In the model, the total number of
units (mass) is conserved, and plays a role of a control parameter.
The distribution function obeys the power-law at a certain critical value.
Below the critical value, the system exhibits an ordinary critical behavior,
and scaling relations hold. Above the threshold, numerical simulations show 
that a power-law distribution coexists with a huge ``monopolized'' member. These 
bahaviors are agreed with the phase transision of the non-conserved process
discussed by Krapivsky and Render\cite{Redner} and Majumdar {\it et al}\cite{Majum}.

\vspace{8mm}
\noindent
{\large\bf 2. Model and Analysis}

The model introduced by Yamamoto {\it et al}, is defined by the following
competition processes and conservation conditions:
\begin{description}
\item{C1.} The system consists of N homogeneous members. The number N
is fixed.
\item{C2.} The system holds $S$ units as resource money. Each member holds
one unit as a minimum amount. The total amount $S$ is conserved.
\item{P1.} Two members, who are picked at random from the whole membership
transact an economic activity competitively. The winner collects the whole
of the competing money and the loser loses all one's money. In order to keep
the number of active members constant, one unit is added to the loser.
\item{P2.} To preserve the total resource money, one unit is reduced from a
member who is picked at random and has resource money more than or equal to
2 units.
\end{description}

We analyze this model by using the z-transform technique and calculate the
probability distribution function $P(X)$ that an arbitrarily chosen member
holds X units of resource money. At steady-states, the equations for
$P(X)$ read
\begin{align}
(1-P(1))(1-P(1)) + \frac{P(2)}{1-P(1)} = &(P(1))^2 \;\;(X=1),\label{11-1} \\
\sum_{i+j=X}P(i)P(j) + \frac{P(X+1)}{1-P(1)}=2P(X) &+ \frac{P(X)}{1-P(1)}
\;\;(X\ge 2).\label{11-g}
\end{align}
Note that $\sum_X P(X)=1$ for normalization. For the analysis of the equation
of this kind, it is useful to consider the z-transform $\phi(z)$ of $P(X)$:
\begin{equation}
\phi(z)=\sum_{X=0}^{\infty}P(X)z^{-X}.\label{ztrans}
\end{equation}
Here it is defined that $P(0)=0$. From the normalization condition, 
$\phi(z)$ must satisfy $\phi(1)=1$. Taking the z-transform of Eq.(\ref{11-g})
and Eq.(\ref{11-1}), we get the basic equations in z-space as,
\begin{equation}
\{1-P(1)\}\phi^2(z)+\{z-3+2P(1)\}\phi(z)-P(1)+\frac1z = 0.\label{11-z}
\end{equation}

\begin{figure}[h]
\begin{center}
\resizebox{6.5cm}{!}{\includegraphics{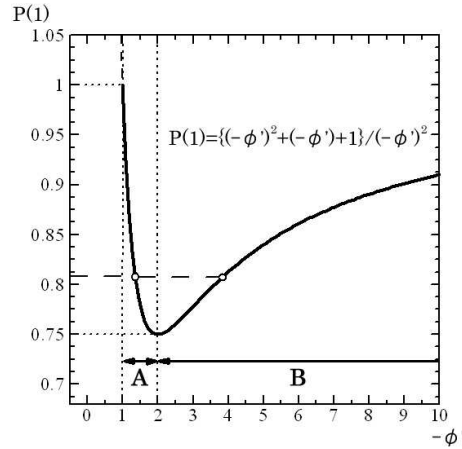}}
\caption{\label{P1}Value of $P(1)$ versus the control parameter
$-\phi'$. $P(1)=1$ means that all the member have one unit.}
\end{center}
\end{figure}

We can now analyze $P(1)$ by differentiation Eq.(\ref{11-z}) with respect to
$z$. Differentiating once and substituting $z=1$, we simply get the identity.
Next differentiation leads to the equation about $\phi'=\phi'(1)$,
\begin{equation}
\{1-P(1)\}\phi'^2+\phi'+1=0. \label{11-p1}
\end{equation}
Note that $-\phi'(1)=\sum XP(X)=S/N$ is the average of resource units, and is
conserved in processes. We employ $-\phi'$ as the control parameter of
the system. Once $-\phi'$ is given, $P(1)$ can be determined via 
Eq.(\ref{11-p1}). Fig.\ ~\ref{P1} shows $P(1)$ versus $-\phi'$. It becomes
evident that $1\ge P(1)\ge \frac34$, and generally $P(1)$ corresponds to
two values $\phi_A$ and $\phi_B$ of $-\phi'$, in the region A:
$1\le\phi_A\le2$ and B: $2<\phi_B$. 
Note that from the definition of the system, each member should have
at least one unit thus $-\phi'=1$ is the minimum value.

First, we treat the case where $-\phi'=2$ and Eq.(\ref{11-p1}) has a 
repeated root, i.e.,  $P(1)=3/4$. Substituting $P(1)=3/4$ to (\ref{11-z}),
we have the solution of the quadratic equation
\begin{equation}
\phi(z)=3-2z\pm2z(1-\frac1z)^{\frac32}.\label{11-phi}
\end{equation}
The singularity of $\phi(z)$ near $z\sim1$ is,
\begin{equation}
\phi(z)|_{z\sim1}\sim1\pm2(1-\frac1z)^{\frac32},
\end{equation}
Then the asymptotic behavior of $P(X)$ at large $X$ is given by
\begin{equation}
P(X)\sim\frac{1}{X^{\sigma}},\;\;\;\sigma=\frac52.
\end{equation}
This power-law exponent coincides with numerical results of ~\cite{Kyamamoto}.

\begin{figure}[h]
\begin{center}
\resizebox{6cm}{!}{\includegraphics{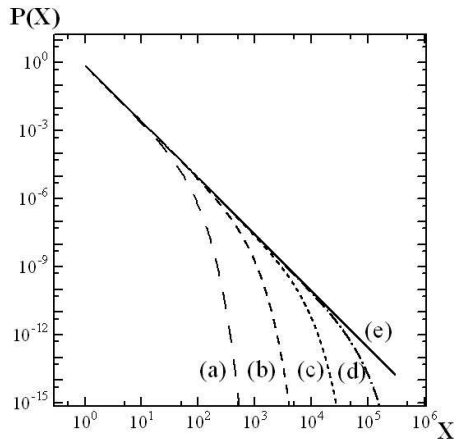}}
\caption{\label{under}Solutions of $P(X)$ with different values of $P(1)$. 
(a) $P(1)=0.76$, (b) $P(1)=0.751$, (c) $P(1)=0.7501$, (d) $P(1)=0.75001$, 
(e) $P(1)=0.75$.}
\end{center}
\end{figure}

Next we discuss the behavior of the system with the control parameter
$-\phi'\ne 2$. The original equations (\ref{11-1}), (\ref{11-g}) can be
solved sequentially if once $P(1)$ is given. In Fig.\ \ref{under},
numerical solutions of Eq.(\ref{11-1}) and (\ref{11-g}) are plotted
for several values of $P(1)$.
With the values $\phi_A$ of $-\phi'$ in the region A of Fig.\ \ref{P1},
our simulation results agree with these sequential solutions. We conclude
that in this region the system has stable steady states, and the solutions
of basic equations (\ref{11-1}) and (\ref{11-g}) represent these stable states.

The behavior of these solutions, which is shown in Fig.\ \ref{under},
is very similar to the ones of ordinary critical phenomena. For arguing
this behavior, let $P(1)=\frac34 + \epsilon$, then analyze the reverse
z-transform,
\begin{equation}
\begin{split}
P(X) &= \frac{1}{2\pi i}\oint z^{X-1}\phi(z)dz \\
&= \frac{1}{2\pi i}\oint z^{X-1}(z-1)\sqrt{1-\frac{1-4\epsilon}{z}},
\label{rev-z}
\end{split}
\end{equation}
in the limit $\epsilon\rightarrow 0$. The most singular part of
(\ref{rev-z}) is given by,
\begin{equation}
\begin{split}
P(X)&\sim \frac{1}{2\pi i}\oint e^{X\{(z-1)-\frac{(z-1)^2}{2}+\cdots\}}
(z-1) \sqrt{z-1+4\epsilon} dz \\
&\sim \frac{1}{2\pi i}\oint e^{X\epsilon \alpha} \epsilon\alpha 
\sqrt{\epsilon\alpha +4\epsilon} \cdot\epsilon d\alpha \\
&\sim X^{-\frac52} \Psi(\epsilon X),
\end{split}
\end{equation}
where the scaling function $\Psi(Y)$ is expressed as,
\begin{equation}
\Psi(Y) = \frac{(Y)^\frac52}{2\pi i}\oint e^{\alpha Y} \alpha
\sqrt{\alpha +4} d\alpha.
\end{equation}
We find that ordinary scaling relations hold.

\begin{figure}[h]
\begin{center}
\resizebox{6cm}{!}{\includegraphics{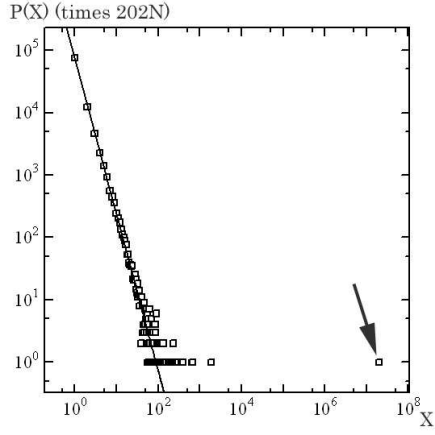}}
\caption{\label{above}Simulation result for $N=100000$ and $-\phi'=202$.}
\end{center}
\end{figure}

The behavior of $P(X)$ is completely different when the control parameter
$-\phi'$ is larger than $2$. Fig.\ \ref{above} shows the simulation result
for $N=100000$ and $-\phi'=202$. The part of units, about $2\times N$ units, 
obey the power-law with the exponent $-5/2$, and the rest of them coagulates
eventually to one ``monopolized'' member, represented by the arrow in Fig.\ \ref{above}.
Thus this member has almost $m_R = (-\phi' -2)\times N$ units.
We have also done simulations with different values of $-\phi'$ in the
range from $3$ to $1002$. In all these simulations $P(X)$ behaves like
Figs.\ \ref{above}. 

The behavior described above is substantially the same as that of the 
model reported by Majumdar {\it et al}\cite{Majum},
though thier model is different from the present model for
it does not preserve the total mass strictly.

\vspace{8mm}
\noindent
{\large\bf 3. Extended Model}

Now we argue that systems which exhibit those behaviors are
not limited to this particular process. To the end ,we examine the model
extended as follows. (i) In the competition process P1, $n$ units are added
to the loser. (ii) In the procedure P2, $m$ member with $X\ge 2$ are picked
at random and one unit is reduced from each of these member. The case 
$n=m=1$ corresponds to the original model. 

Similarly to the original model, the equations for 
$P(X)$ at steady state read
\begin{gather}
\frac{mP(2)}{1-P(1)} = 2P(1) \qquad(X=1),\label{nm-1} \\
\sum_{i+j=n}P(i)P(j)+(1-P(n))(1-P(n))+\frac{mP(n+1)}{1-P(1)}
=(P(n))^2+\frac{mP(n)}{1-P(1)} \qquad(X=n),\label{nm-n} \\
\sum_{i+j=X}P(i)P(j) + \frac{mP(X+1)}{1-P(1)}
=2P(X) + \frac{mP(X)}{1-P(1)} \qquad(X\ne n, X\ge 2).\label{nm-g} \\
\end{gather}

In z-space, these equations are rewritten as
\begin{equation}
A\phi^2(z)+\{m(z-1)-2A\}\phi(z)+m(1-A)\left(\frac1z-1\right)+\frac{A}{z^n}=0, \label{nm-z}
\end{equation}
and the solution of (\ref{nm-z}) is
\begin{gather}
\phi(z)=\frac{1}{2A}\left(-\{m(z-1)-2A\}\pm\sqrt{D(z)}\right), \label{nm-phi}\\
D(z)=\{m(z-1)-2A\}^2-4A\left\{m(1-A)\left(\frac1z-1\right)+\frac{A}{z^n}\right\}. \label{nm-D}
\end{gather}

First we treat the case $n\ne m$. In this case the total amount $S$ of units varies,
thus the situation is quite different from the original model. However, the analysis 
described above is still applicable. 

When $n\ne m$, the discriminant $D(z)$ is
\begin{equation}
D(z)=(z-1)f(z), \qquad f(1)\ne 0. \label{nnem}
\end{equation}
The singularity of $\phi(z)$ near $z\sim 1$ is $\phi(z)|_{z\sim1}\sim1\pm\{(n-m)(z-1)\}^{1/2}$.
It gives a familiar power-law distribution (for example, see \cite{ley, takayasu}),
\begin{equation}
P(X)\sim\frac{1}{X^{\sigma}},\;\;\;\sigma=\frac32.
\end{equation}

When $n<m$, the total amount $S$ eventually vanishes. However if we take $S$ large enough,
our simulation shows the temporal power-law distribution with the exponent $-3/2$. Also in
the case $n>m$, the range of the power-law distribution with the exponent $-3/2$ continually
extends like $\sqrt{t}$ as the time $t$ goes on. 

In the case $n=m$, $-\phi'(1)=-\phi'$ can be used as the control parameter, as the original model,
since $S$ is conserved. When $-\phi'=2$, 
\begin{equation}
P(1)=1-A=\frac{1}{2(n-1)}\left\{2n-\sqrt{2n^2-2n+4}\right\},
\end{equation}
and
\begin{equation}
D(z)=(z-1)^3 g(z), \qquad g(1)\ne 0.
\end{equation}
Thus, the asymptotic behavior of $P(X)$ at large $X$ is given by
\begin{equation}
P(X)\sim\frac{1}{X^{\sigma}},\;\;\;\sigma=\frac52.
\end{equation}
It is just the same exponent as the original model. Also our simulations show that 
the behavior when $-\phi'\ne 2$ is essentially the same as the original. 

\vspace{8mm}
\noindent
{\large\bf 4. Discussion}

These models show drastic change of behavior as the control parameter $-\phi'$ varies.
When $-\phi'>2$, once the ``monopolized'' member appears our simulation shows this
member does not exchange with other members after that. It is practically stable, so we believe that
this transition is not only the normal static phase-transition, but also the dynamic 
``ergodic-nonergodic'' transition. 

Our analysis described above shows that the exponent of ``power-law'' part
does not change in two models. Furthermore, it can be shown analytically 
that the exponent of the power-law depends only on conservation conditions 
when the kernel is constant. These studies suggest strongly that the 
behavior described above is observed universally in conserved aggregation processes. 
We believe this analysis gives clues for classification of these conserved
aggregation processes.

We thank Prof. S. Redner for pointing out references \cite{Redner} and \cite{Majum}
to us.

\end{document}